\documentclass[fleqn,usenatbib]{mnras}

\usepackage{newtxtext,newtxmath}
\usepackage[T1]{fontenc}
\usepackage{ae,aecompl}

\usepackage{graphicx}	% Including figure files
\usepackage{amsmath}	% Advanced maths commands
\usepackage{amssymb}	% Extra maths symbols
\usepackage{multirow}
\usepackage[normalem]{ulem}

\def\N{NGC\,}
\def\NGC{NGC\,}

\def\Ms{$\textrm{M}_{\odot}$}
\def\Ls{$\textrm{L}_{\odot}$}
\def\kms{$\textrm{km~s$^{-1}$}$}
\def\nb{\textsc{NBursts}}

\DeclareMathOperator{\sech}{sech}

\title[A massive thick disc of \N7572]{An excessively massive thick disc of the~enormous edge-on lenticular galaxy \N7572}

\author[A. Kasparova et al.]{Anastasia V. Kasparova,$^{1}$\thanks{Contact e-mail: \href{mailto:anastasya.kasparova@gmail.com}{anastasya.kasparova@gmail.com}}
Ivan Yu. Katkov$^{1,2,3}$ and
Igor V. Chilingarian$^{1,4}$
\\
$^{1}$Sternberg Astronomical Institute, Moscow M.V. Lomonosov State University, Universitetskij pr., 13,  Moscow, 119234, Russia\\
$^{2}$New York University Abu Dhabi, P.O. Box 129188, Abu Dhabi, United Arab Emirates\\
$^{3}$Center for Astro, Particle, and Planetary Physics, New York University Abu Dhabi Abu Dhabi, P.O. Box 129188,\\ Abu Dhabi, United Arab Emirates\\
$^{4}$Smithsonian Astrophysical Observatory, Harvard-Smithsonian Center for Astrophysics, 60 Garden St. MS09, Cambridge, MA 02138 USA\\
}

\date{Accepted 2020 February 27. Received 2020 February 27; in original form 2019 September 22}

\pubyear{2020}

\begin{document}
\label{firstpage}
\pagerange{\pageref{firstpage}--\pageref{lastpage}}
\maketitle

\begin{abstract}
Galactic discs are known to have a complex multilayer structure. 
An in-depth study of the stellar population properties of the thin and thick components can elucidate the formation and evolution of disc galaxies.
Even though thick discs are ubiquitous, their origin is still debated.
Here we probe the thick disc formation scenarios by investigating \N7572, an enormous edge-on galaxy having $R_{25}\approx 25$~kpc and $V_{\rm rot} \approx 370$~{\kms}, which substantially exceeds the Milky Way size and mass. 
We analysed DECaLS archival imaging and found that the disc of \N7572 contains two flaring stellar discs (a thin and a thick disc) with similar radial scales.
We collected deep long-slit spectroscopic data using the 6m Russian BTA telescope and analysed them with a novel technique. 
We first reconstructed a non-parametric stellar line-of-sight velocity distribution along the radius of the galaxy and then fitted it with two kinematic components accounting for the orbital distribution of stars in thin and thick discs.
The old thick disc turned out to be 2.7 times as massive as the intermediate-age thin component, $1.6\times 10^{11}$~{\Ms} vs. $5.9\times10^{10}$~\Ms, which is very unusual.
The different duration of the formation epochs evidenced by the [Mg/Fe] values of +0.3 and +0.15~dex for the thick and thin discs respectively, their kinematics and the mass ratio suggest that in \N7572 we observe a rapidly formed very massive thick disc and an underdeveloped thin disc, whose growth ended prematurely due to the exhaustion of the cold gas likely because of environmental effects.
\end{abstract}


%%%%%%%%%%%%%%%%%%%%%%%%%%%%%%%%%%%%%%%%%%%%%%%%%%

%%%%%%%%%%%%%%%%% BODY OF PAPER %%%%%%%%%%%%%%%%%%

\section{Introduction} \label{intro}

\begin{figure}
\includegraphics[width=\columnwidth]{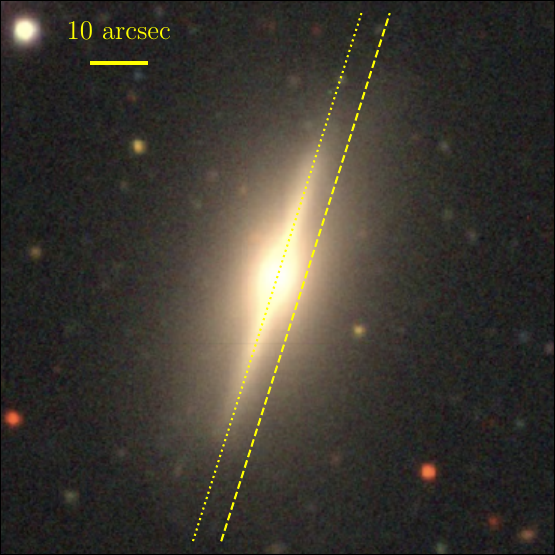}
\caption{DECaLS RGB image of \N7572 retrieved from the web-site \href{http://legacysurvey.org}{legacysurvey.org}.
The dashed and dotted lines show the slit positions for our spectroscopic observations.
The offset between slits is 5~arcsec which corresponds to $\approx4.0$~kpc.}
\label{fig:image}
\end{figure}

We present an in-depth observational study of the edge-on giant disc galaxy \N7572 and the vertical structure of its disc.
\N7572 is a quiescent S0 counterpart of ``super spirals'', the most massive star-forming disc galaxies in the Universe \citep{OLNH16}. 
Its somewhat small diameter compare to the galaxies in \citet{OLNH16} is due to \N7572's large mass-to-light ratio caused by its old stellar populations.
According to our analysis, the \N7572 disc component is five times as massive as the disc of the Milky Way (MW)\footnote{The total mass of the MW stellar disc (thin and thick components) is about $4 \times 10^{10}$~\Ms~\citep{Bland-Hawthorn2016}.}, its rotation velocity reaches 370~{\kms} and the radius is $R_{25}=0.51~\mathrm{arcmin}\approx24.6$~kpc\footnote{Value $R_{25}$ is the 25~mag arcsec$^{-2}$ isophotal radius in the $B$-band.} \citep[HyperLeda,][]{Makarov2014} (see Fig.~\ref{fig:image} and Table~\ref{tab_properties}).
With a total $r$-band luminosity of $1.1\times10^{11}$~L$_{\odot}$ and the corrected colour $g-r=0.80$~mag, \N7572 is as luminous as bright cluster galaxies like Messier~87 in the Virgo cluster and it sits on the bright end of the red sequence populated by the most massive ellipticals and cD galaxies\footnote{The predicted colour at the red sequence for M$_r=-23.0$~mag is $g-r=0.81$~mag \citep{RCSED}.}. 
The unique properties of this object and its edge-on orientation provide us with an opportunity to investigate the vertical structure of the disc components of a galaxy having a significantly larger size than the MW.

This work is a part of our multicycle observational program using the Russian 6-m BTA telescope to study the vertical structure of edge-on galaxies in various environments \citep[see][]{Kasparova16}. 
\citet{White1999} placed \N7572 in the poor cluster WBL\,703 at $z\approx0.044$. However, at the same redshift ($z=0.040$) at a projected distance of $25$~arcmin~$\approx1.3$~Mpc there is the non-virialized cluster Abell\,2572 with two bright merging X-ray cores. 
Hence, this galaxy lives in a fairly dense environment, which can certainly affect its evolution.
We adopt $D=178$~Mpc as a luminosity distance and $D=165$~Mpc as an angular-size distance to \N7572 (the same as the cosmology-corrected luminosity distance to Abell\,2572 from NED). This corresponds to a spatial scale of $\approx 0.8$~kpc~arcsec$^{-1}$ and a distance modulus $m-M=36.25$~mag.

Decades ago \citet{Burstein79} and \citet{Tsikoudi1979} discovered the existence of thick discs during their studies of lenticular galaxies. 
Years later, two coplanar subsystems with drastically different dynamical and stellar population properties were identified in the Milky Way \citep{GilmoreReid1983,Majewski1993, Fuhrmann98, Prochaska2000}.
With the increase of a vertical distance from the mid-plane, the volume density of young metal-rich stars quickly abates and the stellar population becomes dominated by old and metal-poor $\alpha$-enhanced stars \citep[see e.g.,][]{Bland-Hawthorn2016}. 
This phenomenon was attributed to different conditions of formation and/or evolution of the two populations, which were named thin and thick discs \citep{Haywood2013}.
Additional evidences in favour of the different origin of the two disc components in the MW were provided by the modern wide-field imaging and spectroscopic surveys \textit{Gaia} \citep{Gaia2016} and APOGEE \citep{APOGEE2017}.
In particular, \citet{Mackereth2019} demonstrated the very different age--velocity dispersion relations (AVRs) and shapes of the $\sigma_z/\sigma_R$ distributions for stellar populations with low and high [$\alpha$/Fe], corresponding to the thin and thick discs, respectively.
However, there are only a few external galaxies having spectral and photometric data with sufficient quality to clearly separate the stellar population properties of thin and thick discs.

Over the last decade, several teams published results of the analysis of the vertical structure for large samples of edge-on galaxies from photometric data \citep{YoachimDalcanton2006,Comeron11,Comeron2012, Bizyaev2014,Bizyaev2017,Comeron18}.
However, up to now there is less than two dozens of objects with detailed kinematics and stellar population properties of thick discs from spectroscopy \citep{YoachimDalcanton2008,Comeron+2015,Comeron16,Guerou2016,Kasparova16,Fornax3d_2019,Pinna2019_1,Pinna2019_2}. 
Among them, there are no massive galaxies with circular velocities exceeding $300$~{\kms} and only three galaxies have masses comparable to that of the MW.
This severely limits the possibility of comparing the results with the MW disc, and it complicates the verification of the evolutionary scenarios of galactic discs.
It is still unknown, how similar the vertical structure of disc galaxies is for various morphological types from irregular to lenticular and for different luminosities from dwarfs to giants, whether there is a universal scenario of their origin, and how critical the environmental effects are for these processes.
It is worth mentioning that a direct comparison of the properties of the MW's thick disc as defined based on its chemodynamics using individual stars with thick discs in other galaxies defined by photometric analysis is difficult. 
This is because the two definitions do not quite overlap.
One of the reasons for this is that the high-[$\alpha$/Fe] stars (in the Solar neighbourhood in particular) probably have two different origins, the thick disc itself and the inner thin disc regions \citep{Haywood2013,Hayden2017}. 
Nevertheless, we should not completely reject the idea of comparing the multilayer structure of our Galaxy with other galaxies.
Therefore, it is crucial to expand a sample of galaxies with high-quality spectral data for thick and thin stellar populations to be able to identify disc components taking into account both chemistry and kinematics.

There are several scenarios explaining complex vertical disc structure of galaxies. 
Thick discs could be formed as a result of a thin disc secular evolution \citep{Quinn93,Schonrich2009,Villalobos2010}. 
The second possibility is an external origin of thick discs from accreted dwarf satellites \citep[e.g.][]{Abadi2003,YoachimDalcanton2005}. 
The third scenario is a two-stage model implying the thick disc formation at high redshift (z~$\sim 2$) followed by a gradual growth of a thin disc \citep{CMG97,Elmegreen2006, Bournaud2009}.
Spatially resolved internal kinematics of thin and thick discs combined with stellar population analysis can support or debunk a particular scenario in a given galaxy, and this is the way we follow in our study. 

\begin{table}
\centering
\caption{Basic parameters of NGC~7572. 
{
\newline References: 
{[1]} NED (http://ned.ipac.caltech.edu);  
{[2]} LEDA (http://leda.univ-lyon1.fr); 
{[3]} EGIS \citep{Bizyaev2014}. 
We adopted the distance to \N7572 as the cosmology-corrected luminosity distance to Abell\,2572 (from NED, the cosmological parameters: $H_0=67.8$ \kms\,Mpc$^{-1}$, $\Omega_m=0.308$ and $\Omega_{\lambda}=0.692$).
We took $g$ and $r$ magnitudes from EGIS and corrected them for the Galactic extinction (from NED) and $k$-correction \citep{K-corr2010,CZ12}. 
The rotation velocity is from our study.
}}
\label{tab_properties}
 \begin{tabular}{llc}
\hline \hline
RA (J2000.0) & 23$^{\rm h}$16$^{\rm m}$50.$^{\rm s}$368 & -- \\
DEC (J2000.0) & 18\degr28\arcmin59\farcs45 & -- \\
Luminosity distance  & 178~Mpc    &  [1]\\
Angular-size distance   & 165~Mpc    &  [1]\\
Position angle       & 162.3$^{\circ}$ & [2]\\
Rotation velocity    & 368$\pm$8~\kms     & -- \\
$R_{25}$             & 0.51~arcmin  & [2] \\
$M_r$                & $-23.0$~mag  & [3]\\
$g-r$ (corrected)    & 0.80~mag  & [3]\\
\hline
\end{tabular}
\end{table}

The paper is organized as follows. 
Our photometric and spectroscopic analysis is presented in Section~\ref{analysis}; 
in Section~\ref{discuss} and Section~\ref{summary} we discuss and summarize our results; in Appendix~\ref{appendix} we test the stability of our photometric analysis and explore potential pitfalls, which arise when one deals with real ``non-ideal'' galaxies.

\begin{figure*}

\includegraphics[clip,trim={3cm 2cm 2cm 0.75cm},width=0.99\textwidth]{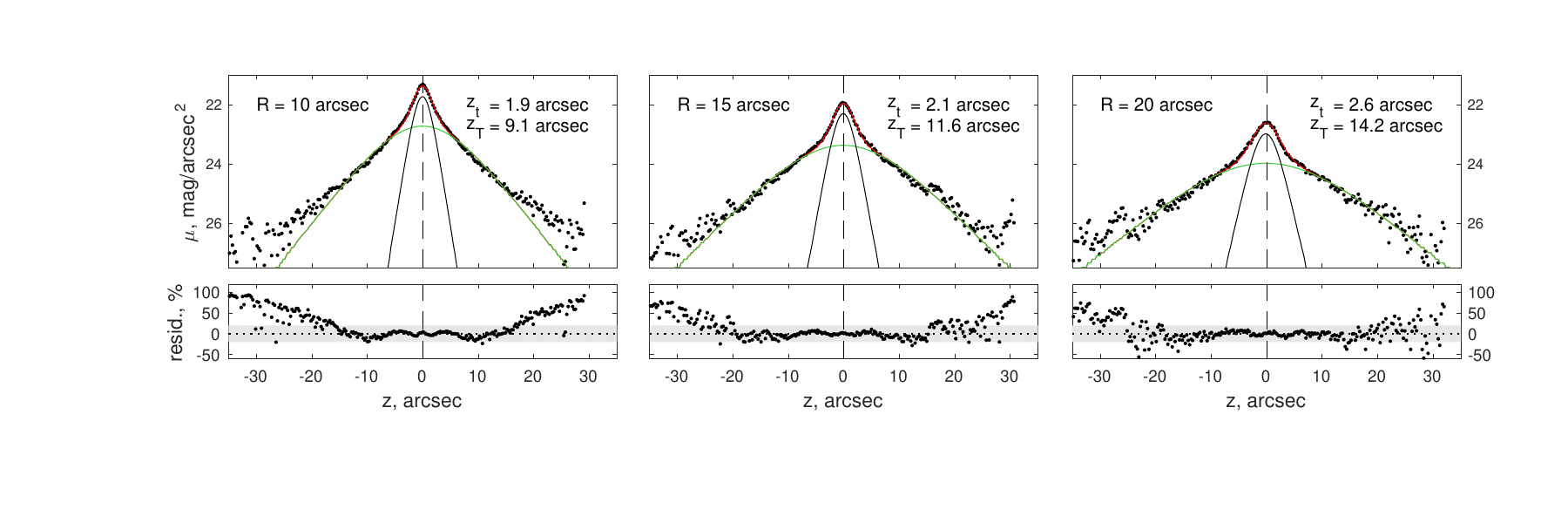}
\caption{The black dots show the vertical observed profiles at $R=10$, $15$, and $20$~arcsec in the DECaLS $g$-band.
The black and green lines mark the contributions of the thin and thick components to the total model (red lines).
The bottom row shows the relative residuals ``(data--model)/data'' and the grey-shaded area covers $\pm20$~per~cent.
 \label{slices}}
\end{figure*}

\section{Analysis of archival data and new observations}
\label{analysis}

\subsection{Photometric analysis} \label{photometry_sect}

We used $g$-band DECaLS images \citep{DECaLS2019} to estimate photometric parameters of the two disc components in \N7572.
We analysed individual one-arcsec wide slices along the vertical axis $z$ (vertical profiles) and obtained the contribution of thin and thick components to the total mass, the relationship between their vertical and radial scales and the vertical scale variations with radius.

\subsubsection{Light profile decomposition}

We used several different techniques for 2D and 1D photometric data analysis while choosing the best method to obtain the disc characteristics of \N7572.
It is often thought that a 2D image decomposition is a more advanced technique than fitting individual slices along the radius. 
However, our analysis of \N7572 images with IMFIT functions for edge-on galaxies of the constant thickness following \citet{Erwin2015} always left strong residuals suggesting the need to take into account the flaring of disc components in the outer part of the galaxy. 
At the same time, adding new functions with \emph{a priori} unknown thickness variations implies a significant increase in the number of free parameters, which lead to the ambiguity and degeneracies between parameters in the best-fitting solution.

Recently, \citet{Comeron18} presented an extensive photometric study of multiple component disc structures in edge-on galaxies \citep[see also][]{Comeron11}.
Their method followed the formalism by \citet{Narayan2002} and implied certain specific assumptions.  
In particular, the scaleheights of the thin and the thick discs are assumed to be constant and their scalelengths similar, 
otherwise the line-of-sight integration in edge-on galaxies becomes non-trivial \citep[see Sec.~3.2.2 in][]{Comeron18}.

We aim to measure thickness variations of the disc components of \N7572, therefore we decided to use a simpler method for the 1D-decomposition of vertical slices at different radial distances $R$, which handles the flaring properly.
In our analysis, we excluded the area of \N7572 affected by the bulge or bar at $|R|<8$~arcsec, which is also dramatically different in stellar velocity dispersion and stellar population properties from the rest of the galaxy (see below).

\citet{Spitzer1942} and later \citet{vdKS81} demonstrated that for an isothermal disc in an equilibrium state, the vertical disc density profiles can be described by the law $I \propto \sech^2(z/z_0)$ where $z_0$ is the scaleheight.
In our work, we use the sum of two such components:
\begin{equation}
\mu_z(R,z) = \mu_t(R) \sech^{2}\left(\frac{z}{z_{t}(R)}\right) + \mu_T(R) \sech^{2}\left(\frac{z}{z_{T}(R)}\right),
\label{eq_sech2}
\end{equation}
where $t$ and $T$ correspond to the thin and thick discs.
Although this approximation is not quite physically justified, it allows us to compare the results with other observational studies and numerical simulations of the vertical structure of discs \citep[and many others]{YoachimDalcanton2006,Bournaud2009,Villalobos2010,Loebman2011}.
However, \citet{Comeron11} and \citet{Comeron2012} compared a more physically motivated fitting method with a double--$\sech^2$ model and demonstrated that it underestimated the ratio between thick and thin discs masses by 20--30~per~cent in total.
Note that for a vertical structure of the MW, the laws of the double--$\exp$ and --$\sech^2$ can be successfully applied \citep[see review by][]{Bland-Hawthorn2016}.

In Appendix~\ref{appendix} we present a series of tests that predict the behaviour of our simplified method to estimate photometric parameters for a widely accepted model of the structure of disc galaxies, two coplanar discs with constant thickness along the radius and different radial scales.
We successfully restore both the vertical and the radial scales of both components.
Also we analyse the reconstruction of disc properties if a galaxy is not viewed exactly edge-on.

\subsubsection{Vertical scales \label{vert_distr}}

In Fig.~\ref{slices} we show three examples of the decomposition of vertical profiles at $R=10$, $15$, and $20$~arcsec.
It is clear that a single-component $\sech^2$--model cannot properly describe the observed profiles.
We also see that at least down to the surface brightness level of 25~mag~arcsec$^{-2}$, that corresponds to about 12~kpc of the vertical distance $z$, the two-component model describes the data quite satisfactory in all three cross-sections along the radius.
However, there is an excess of flux at at 25.5--26~mag arcsec$^{-2}$ and the closer we get to the galaxy centre, the more noticeable the excess light becomes.
It cannot be a stellar halo because it plays an essential role at much lower surface brightnesses in optical bands \citep[see e.g.][]{Peters17}. 
Thus, we probably see the deviation of the double--$\sech^2$ analytical form from observed profiles due to the line-of-sight integration of regions with different thicknesses at different radii because of projection effects or due to the peculiarities of the formation process of a thick disc.
\citet{Qu2011} obtained a similar stellar excess in a series of N-body/SPH models of thick disc formation through minor mergers, which forms due to tidal interactions of the primary galaxy with a satellite happening early during the merger event. 
\citet{Comeron18} found this feature in some most massive galaxies and they interpreted it as a probable sign of the third disc component.

It is worth mentioning the possible effect of the diffuse scattered light. 
According to \citet{Comeron18} and \citet{Stripe82_2019}, the point spread function (PSF) can significantly change the brightness distributions of the disc components, and for edge-on galaxies this effect becomes especially important for vertical profiles.  
Moreover, PSF-related effects can lead not only to increased thickness estimates of the disc components, but also to a counterintuitive decrease in them \citep[see \N0429 in][]{Stripe82_2019}. 
The DECam PSF is substantially narrower than that of the IRAC instrument of the Spitzer Space Telescope and also that of the SDSS. 
In Section~\ref{appendix} we test the possible impact of the DECam PSF on our results. 
Using our method, the vertical scale of the thin disc can be overestimated by 20~per~cent while for the thick disc the effect is no more than 3~per~cent. 
The estimates of the radial scales and the mid-plane surface brightness values do not change significantly.
Also, we cannot explain the excess of light at large heights by PSF-related effects. 
According to \href{https://cdcvs.fnal.gov/redmine/projects/des-sci-verification/wiki/Mapping_of_the_PSF_to_large_radii}{the DECam instrument manual}, the $g$-band PSF at a radius of 10--15~arcsec falls by 12--14~mag~arcsec$^{-2}$ from the maximum. 

\begin{figure}

\includegraphics[clip,trim={0.4cm 1.4cm 1.3cm 0.75cm},width=0.48\textwidth]{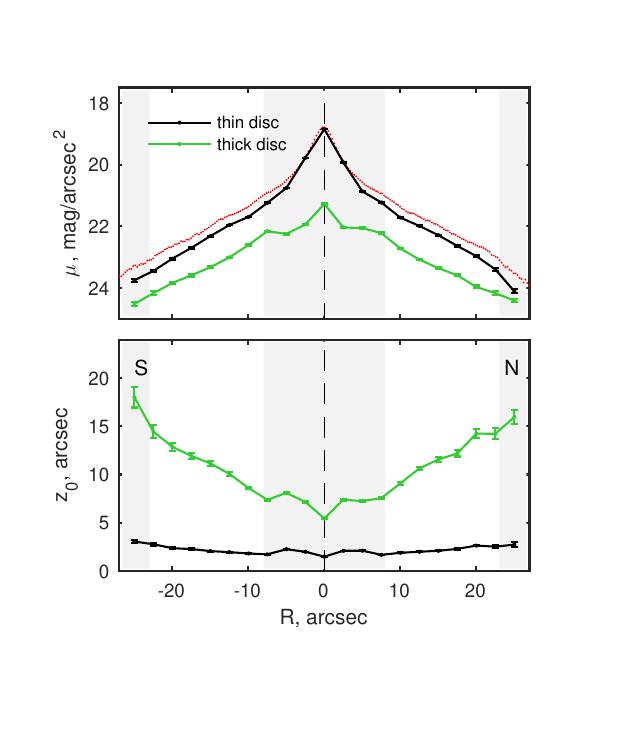}
\caption{The fitted values of the mid-plane surface brightnesses and the vertical scales for the thin (black lines) and thick (green lines) disc components. 
The values are not corrected for the Galactic extinction and $k$-corrections. 
The slightly asymmetric observed axial profile of \N7572 (red dots) has a clear ``knee'' on the Northern side of the galaxy. The grey areas mark the central region affected by the bulge and/or bar and the regions outside the truncation radius at 23~arcsec.\label{vert_scales_and_mu}}
\end{figure}

It is obvious from Fig.~\ref{slices} that the vertical scaleheights of both components increase towards the periphery of the galaxy.
In Fig.~\ref{vert_scales_and_mu} we show the changes of the thickness and mid-plane surface brightness with radius obtained from the decomposition of vertical slices.
Both components have noticeable flaring starting at $8-12$~kpc ($10-15$~arcsec).
The scaleheight of the thin disc in the inner part\footnote{In the Appendix~\ref{appendix} we show that our method satisfactory excludes the bulge influence.} is $z_t=1.9\pm0.1$~arcsec and it grows to $\sim3$~arcsec in the outer regions. 
The scaleheight of the thick component grows from $z_T=7.5\pm0.3$~arcsec by a factor of two.
This proves that the model of the constant disc thickness is not applicable to \N7572.
Here we must keep in mind that choice of the fitting function and the effects of the line-of-sight integration certainly affect the estimate of the disc component thickness. 
In addition, the estimates of $z_t$ and possibly $z_T$ for low-surface brightness regions are a bit overestimated due to the PSF influence.
Therefore, our method provides only a first-order characterization of the flaring.

\subsubsection{True radial scales of the thick and thin discs\label{sect_rad_scales}}

While investigating the photometric properties of edge-on disc galaxies, we certainly want to know the radial scales of the two components. 
In particular, it is important to compare them with the estimates of the corresponding parameters for the MW.
However, the radial photometric profile cannot be unambiguously decomposed into thin and thick components because of (i) the degeneracy between the parameters, and (ii) the lack of \emph{a priori} knowledge of their relative contributions to the total light distribution.

If we measure the exponential scalelength $h$ in an edge-on galaxy fitting the following equation to the integral light profile along the disc major axis:
\begin{equation}
\mu_R\propto\frac{R}{h}K_1\left(\frac{R}{h}\right),
\label{eq_exp_edge-on}
\end{equation}
where $K_1$ is the modified Bessel function \citep{vdKS81}, we will obtain a value for the ``composite'' thin+thick disc.
However, usually this value is taken as a radial scale of the thin disc. 
It is also often assumed, that the thick component radial scale can be derived either from the outer region of the disc beyond break/antitruncation radius of the light profile or at a certain height above the mid-plane \citep[see e.g.][]{Comeron18}. 
However, the ``knees'' on the radial profile of an edge-on galaxy may not be the sign of the transition to the area of thick disc predominance in galactic outskirts but it can be associated with the disc flaring \citep{Borlaff16}. 
And in the case of a non-constant component thicknesses, the thick disc radial scalelength derived from profiles above the mid-plane will become overestimated (see below). 

\begin{figure}
\centering
\includegraphics[clip,trim={0.4cm 1.0cm 1cm 1cm}, width=0.4\textwidth]{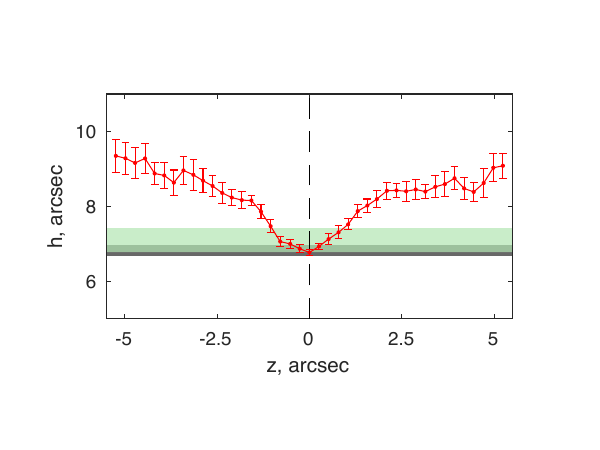}

\caption{The gray and green bands mark $1\sigma$ error ranges of the scalelengths for the thin and thick discs respectively, the red line shows the scalelength of the composite disc at different $z$ as fitted using Eq.~\ref{eq_exp_edge-on}.
\label{true_rad_scales}}
\end{figure}

The light profile of \N7572 exhibits weak ``knees'' at $|R|\approx23$~arcsec better pronounced on the North side (right-hand side from the centre in Fig.~\ref{vert_scales_and_mu}).
There seem to be a thin disc truncation on the right-hand side but within the range $8<R<23$~arcsec both layers are almost purely exponential and have similar scalelengths.
We have derived the scalelengths within this range using Eq.~\ref{eq_exp_edge-on} over the radial profiles of the mid-plane surface brightness of both discs as obtained in our vertical surface brightness fits (Sect.~\ref{vert_distr}).  
In Fig.~\ref{true_rad_scales} the gray and green-shaded areas mark the scalelengths for the thin and thick disc correspondingly, and the red line shows the result of the decomposition of radial profiles of the \textit{composite disc} (total axial profiles including both thin and thick disc light) at different values of $z$.

The scalelengths of the thin and thick components near the mid-plane coincide within uncertainties ($h_t = 6.8\pm0.2$~arcsec and $h_T = 7.0\pm0.3$~arcsec).
The fitted composite disc radial scale above the mid-plane (where a thin component contribution can be neglected) yields a value of the thick disc scalelength only in the ideal case because there can be a complex $h(z)$ trend due to the flaring effect and/or if a galaxy is not seen exactly edge-on (see Appendix~\ref{appendix}).
We can see from the Fig.~\ref{true_rad_scales} that if we try to estimate the radial scale of the thick disc by fitting the surface brightness profiles above the mid-plane with the Eq.~\ref{eq_exp_edge-on} then we always overestimate its value.

\begin{figure}

\includegraphics[clip,trim={0.5cm 2cm 1cm 1cm}, width=0.5\textwidth]{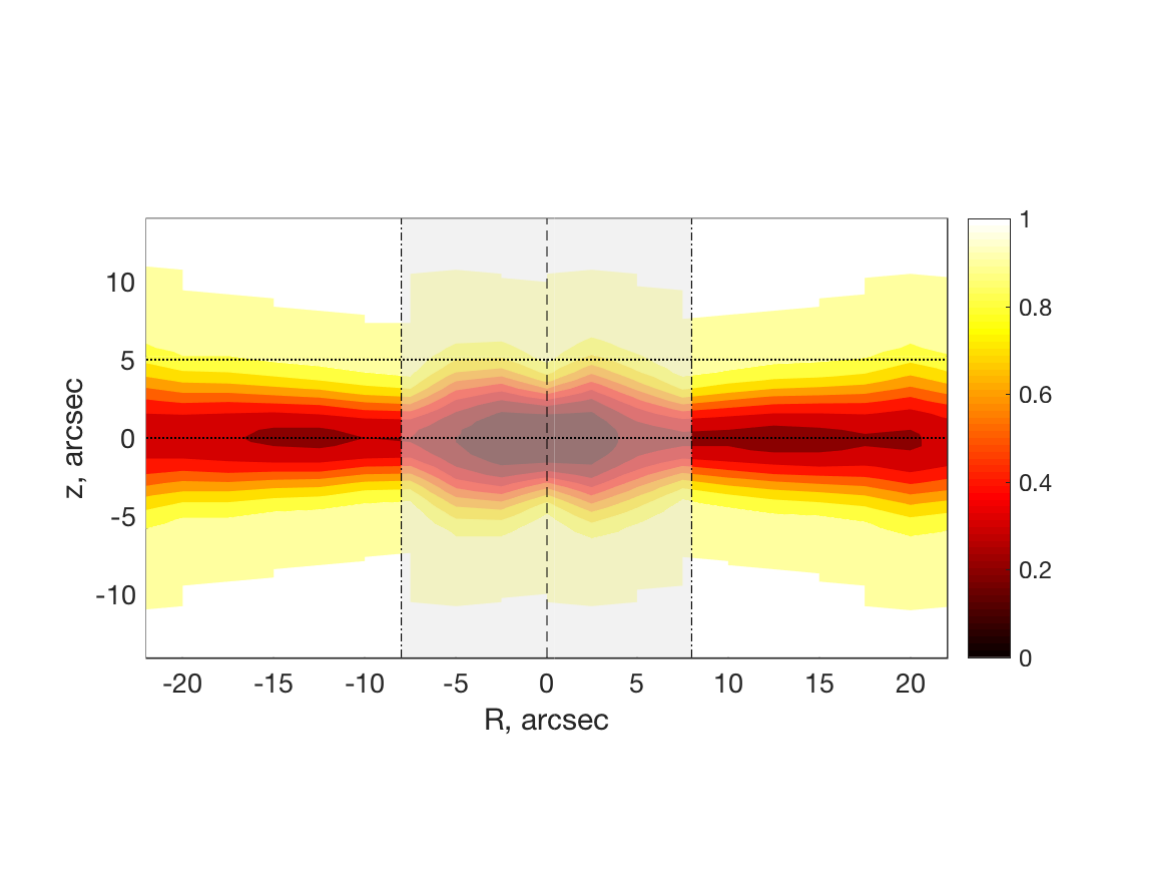}
\caption{The map of thick disc to the total disc (thick $+$ thin components) light ratio.
The area affected by the bulge and/or bar is shaded in grey.
The two dotted lines mark the slit positions for our spectroscopic observations.
\label{ratio}}
\end{figure}

\subsubsection{The relative contributions of the thick and thin discs}
\label{T2t_contribution}

Our decomposition procedure for the vertical profiles described earlier yields, in addition to the scalelengths, the estimate of the thick-to-thin disc ratio over the entire galaxy image.
In Fig.~\ref{ratio} we show a map of the intensity ratio of the thick disc to the total thick $+$ thin disc component.
From Fig.~\ref{vert_scales_and_mu} and Fig.~\ref{ratio} we see that the thick disc contribution in the main plane of \N7572 remains almost constant with radius at about 30~per~cent and it grows up to 90 per~cent when we move away from the mid-plane to $|z|=5$~arcsec, where we placed the slit for our spectroscopic observations (see below).
We can therefore neglect the contribution of thin disc stars at $|z|=5$~arcsec.

Using our photometric analysis we also estimate the total luminosity of the \NGC7572 disc components.
Taking into account the variations of the observed vertical scaleheight with radius, the total luminosity of the thin and thick components in the $g$-band are $1.9\times10^{10}$~\Ls{} and $3.5\times10^{10}$~\Ls{} respectively.
Our estimate of the bulge luminosity from the 2D IMFIT model is about $9.3\times10^{9}$~\Ls.
These values are corrected for the Galactic extinction (from NED) and $k$-correction \citep{K-corr2010,CZ12}.

Our analysis allows us to model what \N7572 would look like seen face-on.
Using our model of the vertical structure of the disc we de-projected it to the face-on orientation and integrated both components in the $z$ direction.
Then we calculated ``face-on'' radial scalelengths, which turned to be $8.3 \pm 0.5$ and $11.4 \pm 2.2$~arcsec for the thin and thick discs correspondingly.
One should note, that these values differ from those in the mid-plane due to the significant flaring. 
In the ``face-on'' view, the dynamically hot old thick disc would contribute over 60~per~cent to the light profiles, while the estimate of the disc radius (four radial scales) becomes about $32$~kpc.
In an enormous disc galaxy like \N7572 seen non-edge-on, we would not be able to directly analyse its thin disc kinematics and stellar populations, because the integrated light would be dominated by its thick disc.

\subsection{Spectroscopic observations and data analysis} \label{sec_spectroscopy}

\begin{figure*}

\includegraphics[width=0.95\textwidth]{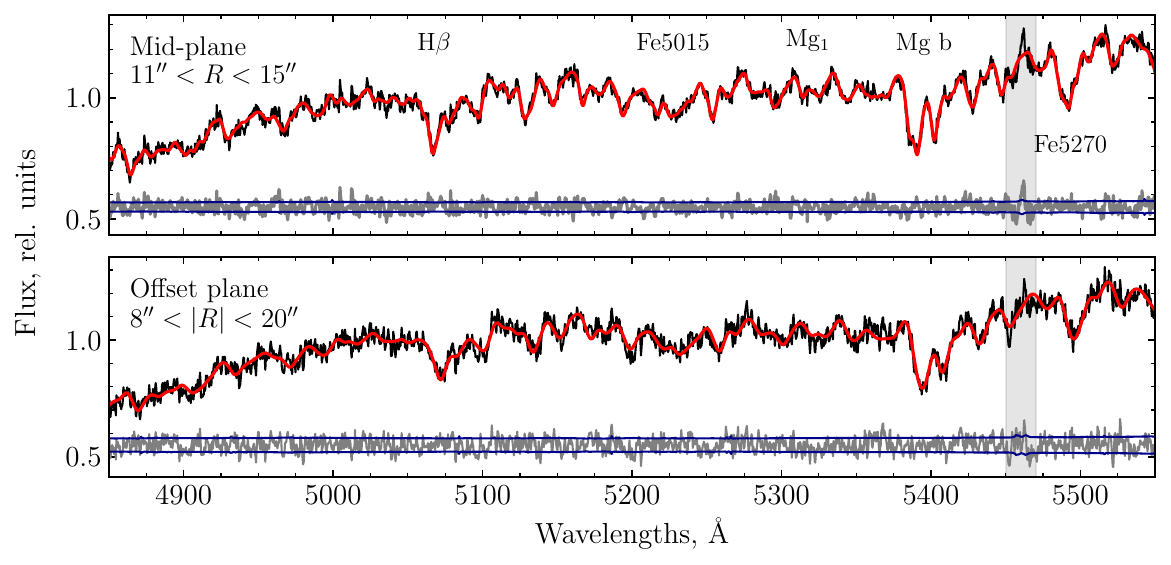}
\caption{Two examples of the full spectrum fitting of \N7572 data sets.
The top panel shows a mid-plane spectrum (in black) for the spatial bin at $R=12.3$~arcsec; the bottom panel corresponds to the total spectrum of the offset slit position co-added in a one large bin ($8<|R|<20$~arcsec).
The red lines show the best-fitting model, residuals (shifted by an arbitrary value) are shown in gray, blue lines represents the Poisson noise propagated through the data reduction steps.
The shaded area shows masked region around the 5460~\AA\ night sky emission line.
}
\label{fig:spectra_fits}
\end{figure*}

This study is a continuation of our observational program for which we described the methodology and presented first results of spectroscopic data analysis in \citet{Kasparova16}.
To estimate the stellar population properties of the thin and thick disc components of \N7572 we collected deep long-slit spectra using two slit positions (see Fig.~\ref{fig:image}), which cover the galaxy mid-plane and the area above the mid-plane at $z=5$~arcsec, where the thick disc prevails.

\subsubsection{Observations and data reduction}

We observed \N7572 using the universal spectrograph SCORPIO \citep{Afanasiev2005} operated at the prime focus of the Russian 6m BTA telescope.
We used the volume phase holographic grism ``VPHG2300G'' which provides a spectral resolution of full width at half-maximum FWHM $=2.2$~\AA\ corresponding to $\sigma_\mathrm{inst}\approx55$~\kms\ in terms of velocity dispersion with a 1~arcsec wide 6~arcmin long slit in the wavelength range $4800-5600$~\AA.
We have obtained both mid-plane and thick disc spectra on the night of 07--08/09/2016 under good atmospheric conditions (seeing FWHM of 1.4~arcsec) with the exposure times of 6000~s (mid-plane) and 9000~s (thick disc).
The slit positions are shown in Fig.~\ref{fig:image}.
The CCD EEV42-40 detector ($2048\times2048$ pixels) provided a spectral sampling of 0.37~\AA~pix$^{-1}$ and a slit plate scale of 0.36~arcsec~pix$^{-1}$ with the $1\times2$ binning.
In addition to the science spectra, we obtained night-time internal flat fields and He-Ne-Ar arcs, and also twilight spectra and a spectrophotometric standard star.

We reduced the spectroscopic data with our own {\sc idl}-based pipeline that included the following steps: bias subtraction, flat fielding, cosmic ray hit removal using the Laplacian filtering technique \citep{lacosmic}, wavelength calibration and linearization, sky subtraction and flux calibration.
SCORPIO has an instrumental line spread function (LSF) with a complex non-Gaussian shape, which varies both along and across the dispersion directions.
To account for LSF variations in the subsequent analysis, we determined its shape with a Gauss--Hermite parametrization by using high signal-to-noise twilight spectra observed with the same instrumental setup.
We have estimated the night sky background from the outer slit regions not covered by our target galaxy and then used an optimized sky subtraction technique that takes into account LSF variations along the slit \citep{skysubtr_adass_proc2011,katkov2014_ilgpop_skysubtr}.
We computed flux uncertainties from the photon statistics at the initial stage of the reduction and propagated them through all reduction steps.

\subsubsection{Full spectral fitting}
\label{subsec:spectral_fitting}

To derive internal kinematics and stellar population properties (mean ages, metallicities [Fe/H] and $\alpha$-elements abundances [Mg/Fe]) of thin and thick discs, we applied the \nb\ full spectrum fitting technique \citep{nbursts_a,nbursts_b}, which we updated by including the fitting of $\alpha$-element abundances.
We used two different grids of simple stellar population (SSP) models.
To better constrain internal kinematics we used the high-resolution ($R=10000$) \textsc{pegase.hr} SSPs \citep{pegasehr}.
Age, metallicity, and $\alpha$-abundance of stellar populations were derived using intermediate-resolution ($R\approx2000$) {\sc miles v11} models \citep{miles2015_alpha} where two SSP grids are available for  solar-scaled [Mg/Fe]\,$=0.0$~dex and $\alpha$-enhanced [Mg/Fe]\,$=+0.4$~dex populations with the \citet{Kroupa02} stellar  mass function.

The \nb\ technique implements a pixel space $\chi^2$ minimization algorithm where an observed spectrum is approximated by a stellar population model broadened with a parametric line-of-sight velocity distribution (LOSVD) and multiplied by a polynomial continuum to take into account dust attenuation and/or possible flux calibration imperfections in both observations and models.
The principal computational step during the $\chi^2$ minimisation is the spectrum interpolation from the stellar population grid according to the given model parameters.
For the \textsc{pegase.hr} models we used 2D spline interpolation across ages and metallicities implemented in the original version of \nb.
The {\sc miles v11} library has two values of [Mg/Fe] (0.0~dex and +0.4~dex).
To include the [Mg/Fe] dimension use this grid we modified the interpolation procedure by running the 2D spline interpolation twice for both [Mg/Fe] values and then linearly interpolating along the [Mg/Fe] axis.
This approach provided fully consistent results with the direct scan of the $\chi^2$ space over an oversampled [Mg/Fe] axes presented in \citet{CA18}.
To take into account the instrumental broadening, we pre-convolved the grid of stellar population models with the LSF derived from twilight spectra as described earlier.

Prior to the full spectrum fitting, we binned our long slit spectra in the spatial direction using bins with sizes linearly increasing when moving away from the galaxy centre and simultaneously requiring a certain minimal signal-to-noise ($S/N$) ratio\footnote{We computed the $S/N$ in the spectral range of $5270\pm30$~\AA.} ($(S/N)_{min}=16$ and 11 in cases of the mid-plane and offset spectra respectively).
The binning along the slit differs for kinematic and stellar population profiles in both spectroscopic data sets. 
Our spectra in the offset slit position that covered the thick disc component have a relatively low $S/N$.
Therefore for stellar population analysis, we computed a large spatial bin ($8<|R|<20$~arcsec) in addition to the regular bins.
Before co-adding the spectra within this bin, we accounted for the galaxy rotation curve shifting each spectrum at given unbinned pixel to the same systematic velocity. 
Examples of spectra with their best-fitting models are shown in Fig.~\ref{fig:spectra_fits}.
The mid-plane spectrum has been extracted at a radius of 12.3~arcsec, and the thick disc spectrum is binned over the whole large bin.
We estimated uncertainties of the derived parameters by running a Monte-Carlo simulation for a hundred realization of synthetic spectra for each spatial bin, which were created by adding a random noise to the best-fitting model corresponding to the signal-to-noise ratio in that bin.

\begin{figure*}

\includegraphics[width=0.95\textwidth]{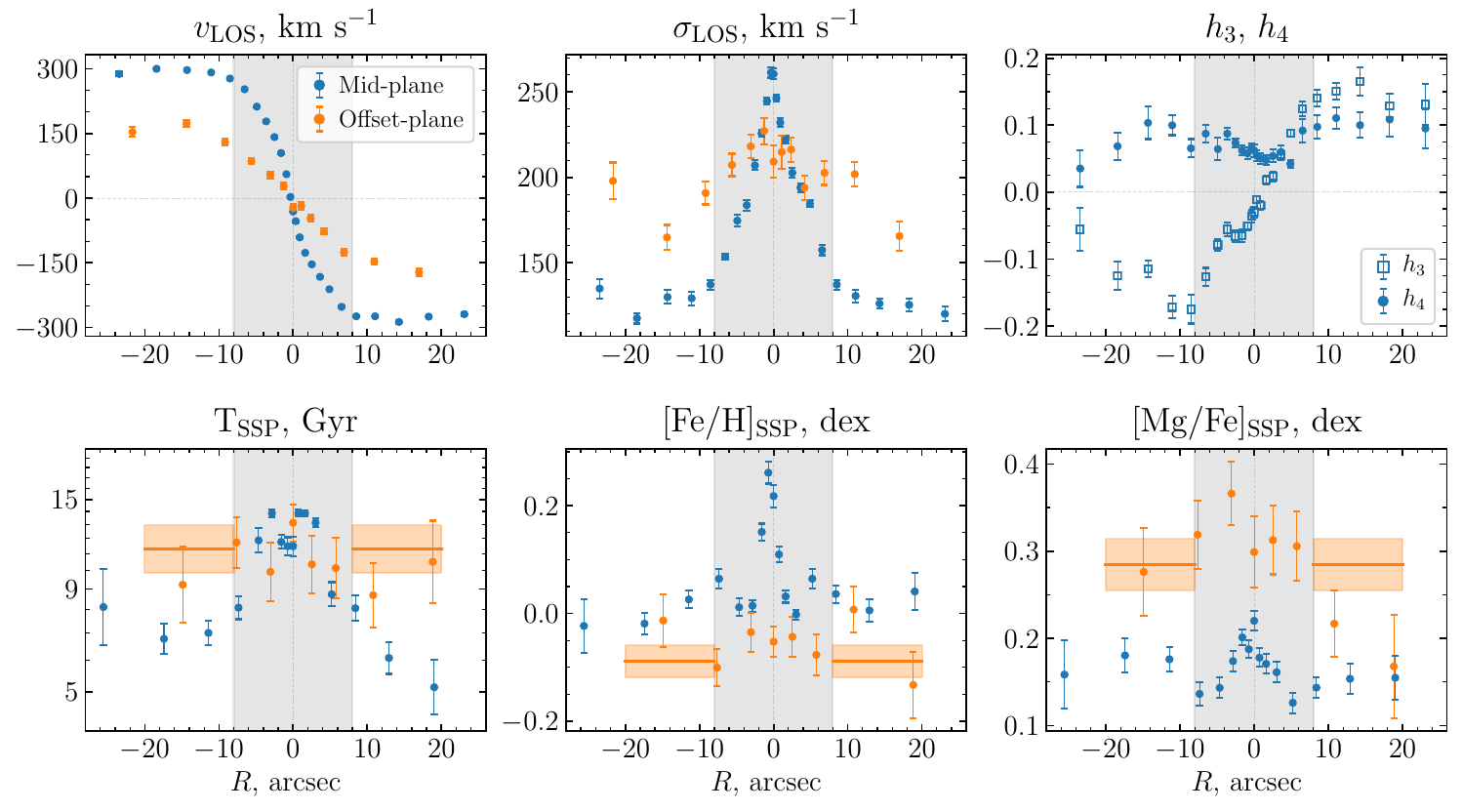}
\caption{Radial profiles of stellar population properties derived from the SCORPIO long-slit spectra.
The blue and orange colors correspond to the mid-plane and offset slit positions, respectively.
We did not find significant deviations from a Gaussian LOSVD for offset spectra, hence we show $h_3$, $h_4$ coefficients for the mid-plane spectra only.
Note that the kinematics and stellar population parameters are computed for different spatial binning layouts (see Sec.~\ref{subsec:spectral_fitting}).
The offset spectrum had a relatively low S/N, therefore we measured stellar populations in one large bin shown by the orange line. 
The shaded orange area around the line corresponds to the uncertainties for the stellar population parameters in that bin.
The grey shaded area shows the region where the bulge or bar has a significant contribution to the integrated light.}
\label{st_pop}
\end{figure*}

The reconstructed stellar population properties for both slits are shown in Fig.~\ref{st_pop}.
We see a clear difference in the parameters between the mid-plane and offset spectra where, respectively, the thin and thick discs dominate.
Outside the bulge area ($|R|>8$~arcsec), we see an intermediate age stellar population $T_{\rm SSP}\approx5\dots8$~Gyr for the mid-plane, and an older $10\dots13$~Gyr population for the offset-plane.
Metallicities are solar or slightly sub-solar in the disc regions in both slits.
The $\alpha$-abundance values exhibit a clear difference between the mid-plane [Mg/Fe]\,$\approx0.15$~dex and offset slit positions [Mg/Fe]\,$\approx0.3$~dex.
We do not find significant radial gradients of the stellar populations properties of the disc components.

The central area of the galaxy has an old stellar population $T_{\rm SSP}\approx 12$~Gyr. 
This is similar to the age of the thick disc which dominates the contribution to the offset slit spectrum even close to the principal axis of the galaxy.
There is a sharp metallicity peak in the nuclear part of the galaxy, where [Fe/H] reaches 0.25~dex compared to 0.05~dex a few arcsec away, which is a sign of a chemically decoupled nucleus.
From the available data, we cannot tell whether we see a real bulge in the central region of \N7572 or it is a huge bar in its thin disc.
The latter option is favoured by relatively low [Mg/Fe] values, however, whether a bar can exist within such a dynamically hot disc is a question that can be answered only with dedicated numerical simulations.
The line-of-sight velocity in the mid-plane reaches the plateau at about 300~\kms, the velocity dispersion in the disc dominated area is about 120~\kms.
In the offset-plane, $v_{\rm LOS}$ reaches 150~\kms\ and the velocity dispersion is as high as $\sigma_{\rm LOS}\approx180$~\kms.

From the stellar population properties estimated from {\sc miles v11} SSP models we have calculated the average mass-to-light ratios $(M/L)_{t}=3.1$ and $(M/L)_{T}=4.6$ in Solar units for the thin and thick components in the SDSS $g$-band.

The spectral resolution and high signal-to-noise ratio of the data allows us to estimate the coefficients $h_3$ and $h_4$ along the mid-plane slit.
The non-zero values of $h_4$ and, in particular, $h_3$ coefficients indicate strongly non-Gaussian and asymmetric LOSVD shape.
An anticorrelation of velocities and $h_3$ profiles (as in the case \N7572) is known for edge-on galaxies \citep{ChungBureau2004}, but such large $h_3$ values (up to 0.2) are suggestive that it could be the result of a noticeable contribution of the thick disc component with a low rotation velocity compared to that of the thin disc.
We did not find significant deviations from the Gaussian LOSVD shape in the offset spectra as evidenced by $h_3$ and $h_4$ having zero values within uncertainties.

\subsubsection{A non-parametric LOSVD reconstruction and a two-component model of disc kinematics}
\label{sec:nonpar_losvd}

\begin{figure*}

\includegraphics[width=0.95\textwidth]{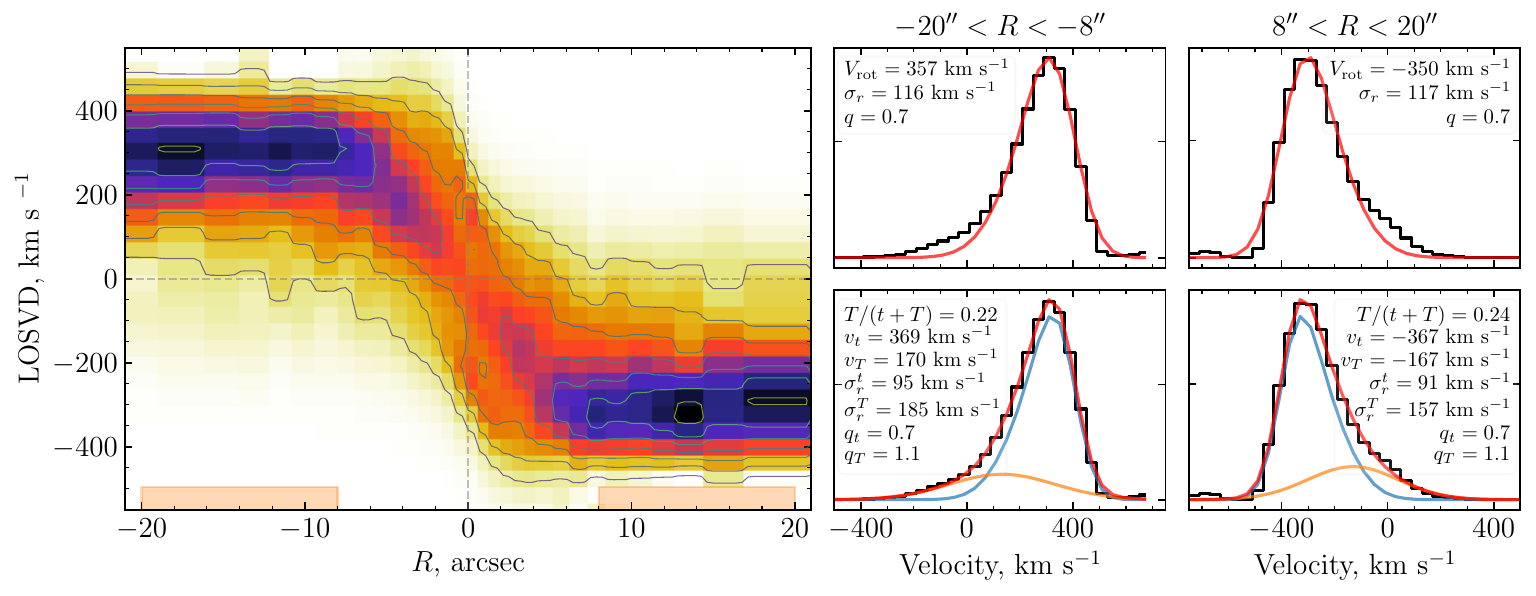}
\caption{The left-hand panel shows a non-parametrically reconstructed stellar LOSVD along the mid-plane slit. 
The right-hand panel demonstrates the cross-sections of the stellar LOSVD as a black stepped line for the two radial bins ($8<|R|<20$~arcsec) in the disc-dominated region of \N7572.
The red lines on the two top right-hand panels show a one-component model of the LOSVD obtained as a result of the line-of-sight integration of the luminosity-weighted Schwarzschild orbital distribution in the disc (see details in Section~\ref{sec:nonpar_losvd}).
Bottom right-hand panels show the result of the two-component modelling where blue and orange lines correspond to the contribution of the thin and thick discs to the total LOSVD.}
\label{fig_losvd}
\end{figure*}

To clarify whether the thick disc influences the mid-plane kinematics, we reconstructed the LOSVD in a non-parametric way.
The deconvolution can be considered as a linear inverse problem whose solution $\mathcal{L}=\{l_1,...,l_n\}$ can be estimated using a least-squares technique.
We used a zero-order regularization \citep[Sec.~19.4.1,][]{num_recipes} to stabilize a solution, which is very sensitive to noise in the data.
The regularized inverse problem can be expressed as:
\begin{equation}
    \min_\mathcal{L} ||Y - A\cdot\mathcal{L}||^2 + \lambda ||\mathcal{L}||^2,
\end{equation}
where $Y$ is an observed spectrum logarithmically rebinned in the wavelength domain, $A$ is a $n\times m$ matrix, where each row contains a template spectrum of the length of $m$ pixels shifted in velocity corresponding to a given $l_i$ element of the LOSVD $\mathcal{L}$, $\lambda$ is a free regularization parameter.
To solve this problem numerically, we used the bounded-variable least-squares (BVLS) method \citep{bvls} implemented by M.~Cappellari\footnote{The BVLS routine implemented by  M.~Cappellari is available on his web page \url{http://www-astro.physics.ox.ac.uk/~mxc/software/}} because we need to impose the positivity constraint on the LOSVD.
As a template, we used the best-fitting SSP models from the \nb\ stellar population analysis which we broadened only with the instrumental LSF.
We applied a similar approach in the past to the LOSVD reconstruction in several galaxies with counter-rotating stellar discs \citep{Katkov2013ApJ_ic719, Katkov2016MNRAS_n448}.
Fig.~\ref{fig_losvd} demonstrates the result of the stellar LOSVD reconstruction in the mid-plane spectra for the two spatial bins symmetrically placed on the different sides of the galaxy at $-20<R<-8$ and $8<R<20$~arcsec.

The reconstructed stellar LOSVD looks very asymmetric with extended tails towards low rotational velocities (see Fig.~\ref{fig_losvd}) which we attribute to the presence of a slowly rotating component originating from the thick disc.
To quantify the thick disc contribution to the line-of-sight integrated spectrum we first applied a simplified approach by decomposing the reconstructed LOSVD into two Gaussian components assuming a narrow high-velocity component corresponding to the thin disc and a broad slowly rotating component corresponding to the thick disc.
The resulting best-fitting model did not look reliable, because it yielded almost equal contributions of the two components, which was in clear contradiction to the photometric profiles decomposition (see above).

Then we applied a physically motivated approach by integrating the luminosity-weighted LOSVD along the line-of-sight and taking into account the orbital structure of disc components. 
Namely we assumed that (i)~the LOSVD for a given point along the line-of-sight is a projection of the Schwarzschild orbital distribution (a set of Gaussian distributions along each coordinate in velocity space) at a given galactocentric distance; (ii) the weights of the LOSVD exponentially decline with galactocentric distance with scales derived from the photometric analysis; (iii) there is no gradient of radial and azimuthal velocity dispersions.
The parameters of our model are: the circular velocity $V_{\rm rot}$, the radial velocity dispersion $\sigma_r$, and the ratio of the radial to azimuthal velocity dispersion components $q=\sigma_\phi / \sigma_r$.
Our technique is in fact an application of the approach described in \citet{ZasovKhoperskov2003} to observational data.
Because of the strong degeneracy between $\sigma_r$ and $q$ we fixed $q$ to the value of $1/\sqrt{2}$ calculated for the flat rotation curve under the epicyclic approximation, which should be valid for a relatively dynamically cold thin disc.
The two top panels on the right half of Fig.~\ref{fig_losvd} clearly show that the reconstructed LOSVDs (black stepped lines) are poorly fitted by a single-component model (red solid lines).

Then we used a two-component model where $q$ was fixed for the thin disc component and was left as a free parameter for the thick disc, which is assumed to be too hot for the epicyclic approximation to be valid.
A two-component model has one additional parameter, a relative contribution of the components.
In both radial bins the reconstructed LOSVDs are very well reproduced by the model (see bottom panels on right-hand side of Fig.~\ref{fig_losvd}). 
From this modelling, we obtained the radial velocity dispersion of the thin disc of $\sigma_r^T=93\pm3$~\kms{} and a much larger value for the thick disc $\sigma_r^T=171\pm20$~\kms.
The ratio between the dispersion components of the thick disc $q\approx 1$ corresponds to the isotropic stellar velocity dispersion distribution.
The corrected thick disc rotation velocity turns out to be $\approx 170$~\kms\ while for the thin disc it reaches $\approx 370$ \kms.
Note that the rotational velocity estimates from our modelling are free from the asymmetric drift effect and represent the true circular velocity of a galaxy at a given distance from the galaxy centre.
The two-component LOSVD decomposition yields the contribution of the thick disc to the mid-plane spectra of 23~per~cent, which is fully consistent with the photometric analysis ($\approx30$ per~cent, see Sect.~\ref{T2t_contribution}).

\begin{table}
\centering
\caption{The observed properties of the thin and thick disc components of \N7572. In two last blocks of the table corresponding to the results of the spectral analysis we provide the approximate averaged values for the radial range $8<|R|<20$~arcsec.}
\label{tab_res}
 \begin{tabular}{lll}
\hline \hline
               & Thin disc  & Thick disc \\
\hline
\multicolumn{3}{l}{Photometry}\\
\hline
Radial scalelength  & $5.4\pm0.1$~kpc & $5.6\pm0.3$~kpc\\
Central scaleheight & $1.5\pm0.1$~kpc & $6.0\pm0.2$~kpc\\
Flaring        & Yes & Yes\\
Total mass & $5.9\times10^{10}$~\Ms & $1.6\times10^{11}$~\Ms\\

\hline
Slit positions: & $z=0$~kpc & $z=4$~kpc\\
\hline
$v_{\rm LOS}$ &$300$~km~s$^{-1}$&$150$~km~s$^{-1}$\\
$\sigma_{\rm LOS}$&$120$~km~s$^{-1}$&$180$~km~s$^{-1}$\\
$T_{\rm SSP}$        & 5--8~Gyr & 10--13~Gyr\\
\textrm{[Fe/H]}         & $0.0$~dex & $-0.1$~dex\\
\textrm{[Mg/Fe]}        & 0.15 & 0.3 \\
$(M/L)_g/(M/L)_{\odot,g}$        & 3.1 & 4.6 \\
\hline
\multicolumn{3}{l}{LOSVD analysis of the mid-plane spectrum}\\
\hline
Rotation velocity $V_{\rm rot}$ & $370$~\kms & $170$~\kms \\
Radial dispersion $\left<\sigma_r\right>$ & $90$~\kms & $170$~\kms \\
$q=\sigma_\phi / \sigma_r$ & 0.7 & $\approx1$\\

\hline
\end{tabular}
\end{table}

\section{Discussion}
\label{discuss}

There are several scenarios explaining the difference of stellar population properties of thin and thick discs. 
(i) A thick disc could have formed as a result of the thin disc secular evolution that changed its radial and vertical structure by the effects of the radial migration \citep{Schonrich2009, Minchev2010, Loebman2011,Minchev2015}, the dynamical heating by giant molecular clouds \citep{Spitzer1953,Lacey1984} and/or minor mergers
\citep{Quinn93,Villalobos2010,Qu+2011b}. 
(ii) An accretion scenario implies that stars which we observe as a thick disc were predominantly formed in dwarf satellite galaxies which merged with the host and built up a dynamically heated discy component \citep[e.g.][]{Abadi2003}. 
(iii) A two-stage model describes the thick disc formation at high redshift (z~$\sim 2$) in a high-density and high velocity dispersion interstellar medium \citep{Elmegreen2006, Bournaud2009,Lehnert2014}.
After that, a thin disc is formed from an additional supply of gas from an external source. 
The new thin gas layer may come from intergalactic filaments \citep{CMG97, Combes14}, minor wet mergers \citep{Silchenko2011NGC7217} or as a result of the cooling of gas ejected earlier through stellar feedback \citep{Fraternali2013ApJ} or left from the first (thick) disc formation phase \citep[the ``upside-down'' models by][]{Burkert1992,Bird2013}.
The latter group of scenarios naturally explains the presence of an observed time delay between the onset of star formation in the thin and thick discs of the Milky Way \citep[see e.g.][]{Lehnert2014,Kilic2017}. 

A thick disc formation in a major accretion event is not likely because there is a too low fraction of counter-rotating stars in thick discs of massive galaxies, which should be observed more commonly if such scenario took place \citep{YoachimDalcanton2008_kin, Comeron+2015, Comeron19, Kasparova16}. However, for the definitive rejection of this scenario it is critical to obtain observational data on the vertical structure of giant discs. 

Overmassive giant disc galaxies are very rare and poorly studied \citep[see e.g.][]{Saburova2018}. 
The fact that we obtained high-quality spectrophotometric data for \N7572, an edge-on galaxy with a size comparable to those discussed in \citet{Saburova2018}, is therefore important.
According to \citet{Saburova2018}, giant galaxies with $R_{25}>30$~kpc represent about 1~per~cent of the total population of discs with inclination angles $i>40^{\circ}$. 
The objects with rotation velocities $>350$~{\kms} are 4~per~cent of the Saburova's sample of giant galaxies. 
According to modern concepts, nearly all giant galaxies with masses $>10^{12}$~{\Ms} should have experienced at least one major merger \citep{Rodriguez-Gomez2015}. 
During such an event the initial disc structure is thought to be destroyed but a new disc could reform later in some cases \citep{Puech2012,Peschken2019}. 
The principal question is ``what stellar population properties and vertical structure peculiarities of such a newly formed disc are?''

Each thick disc formation mechanism leaves a fossil record in the observed kinematics and stellar populations properties of galaxies.
Below we discuss the key observable facts regarding \N7572 (Table~\ref{tab_res}) which we compare with the Milky Way and other objects.
Then we use them to test different scenarios for the thick disc origin.
We compare our data with numerical models, which were made mainly for MW-like galaxies in terms of mass and size.
It would be of interest to have libraries of galaxy simulations covering very massive galaxies such as the one studied in here.

\subsection{The observed properties of \N7572}

\N7572 is a giant \textit{lenticular} galaxy with no signs of the cold gas presence or current star formation, which is confirmed by the absence of emission lines in the mid-plane spectrum and the H{\sc i} non-detection according to \citet{Springob2005}.
Its rotation velocity reaches $370$~{\kms} and the total disc mass is more than five times the mass of the MW disc. 
Such high rotation velocities are typical for ``super spirals'' \citep{OLNH16} and giant LSB galaxies \citep{u1922} and they are extremely rare.
\N7572 has complex vertical structure and to properly describe it,  we need at least two (thin and thick) flaring components with very similar mid-plane radial scales.
The thick disc stellar mass is more than double the thin disc mass, $1.6 \times 10^{11}$~\Ms{} versus $5.9 \times 10^{10}$~\Ms.
The thick component contribution to the total flux increases from 30~per~cent to 90~per~cent when we move from the mid-plane to $|z|=4$~kpc above and it slightly varies with the radius. 

We have not detected any significant radial gradients of the stellar population properties outside the bulge area.
In the mid-plane we see intermediate age stars ($t_{t}=5\dots8$~Gyr) while the stellar population of the thick disc is older $t_{T}=10\dots 13$~Gyr.
Despite the similarity in metallicities between the thin and thick discs ([Fe/H]$_{t}\approx 0$~dex and [Fe/H]$_{T}\approx-0.1$~dex) the difference in the $\alpha$-abundance is statistically significant ([Mg/Fe]$_{t}$=+0.15~dex vs [Mg/Fe]$_{T}$=+0.3~dex).
This suggests that the formation epochs for the two disc components had different duration \citep[see e.g.][]{Thomas+05}. 
According to Eq.~4 in \citet{Thomas+05} the thick disc was formed quickly in $\sim0.25$~Gyr while the thin disc grew over a much longer period of time (2~Gyr).

From the non-parametric modelling of the stellar LOSVD for the mid-plane spectra we estimated the rotation velocities and radial velocity dispersions of the thin and thick discs in the galactic main plane: $\upsilon_{t} \approx 370$~\kms{} and $\langle\sigma_r^t\rangle\approx90$~\kms\ for the thin disc and $\upsilon_{T} \approx 170$~\kms{} and $\langle\sigma_r^T\rangle\approx170$~\kms\ for the thick disc.
The high value of $\upsilon_{T}/\sigma_r^T$ argues that the thick disc of \N7572 is a dynamically hot subsystem where the epicyclic approximation is invalid and, therefore $q=\sigma_{\phi}/\sigma_r$ must be treated as a free parameter in the LOSVD modelling.
For the thick disc we derived $q\approx1$ while for the thin component we fixed it to $q=1/\sqrt{2}$ following the epicyclic approximation, which is valid for a relatively cold thin disc ($\sigma_r \ll \upsilon$).
Assuming that $q$ does not vary along the $z$ axis and accounting that thick disc line-of-sight velocity dispersion at 4~kpc above the mid-plane is very close to that in the mid-plane value, we conclude that $\sigma_{r}(z)$ for the thick disc of \N7572 is close to constant  \citep[similar to the MW according to][]{Mackereth2019}.

We detected extended wings of the vertical photometric profiles noticeable in the inner part of the disc (the light excess is found at $|z| > 15$~arcsec) not described by the two-component model.
\citet{Comeron18} have found similar features in 6~per~cent of their sample (and it is precisely the most massive galaxies) despite using the self-consistent equations to define the vertical profiles and taking into account the diffuse scattered light.
\citet{Qu2011} obtained a similar additional stellar excess in the numerical models of thick disc formation through minor mergers.  
Although this requires numerical confirmation, we believe that this effect can be explained in two more ways:
(i) the line-of-sight integration of the light from two flaring and not exactly $\sech^2$--like disc components should yield a flux excess, which will be more noticeable at small $R$, as we observe in \N7572;
(ii)~the wings might represent an additional structural component of the disc if we consider a two-stage formation scenario of the vertical disc structure \citep[e.g.][]{CMG97} as the simplest case of a multistage scenario, which can, in principle, lead to the formation of three or more disc layers \citep[see also \N4013 in][]{Comeron2011_N4013}.
The likelihood of such repeated disc formation episodes should be higher for massive galaxies.
The two main questions to be answered regarding this scenario are ``how much time is needed for each disc formation stage?'' and ``do we have the instrumental sensitivity to obtain observational confirmation of this scenario?''

\subsection{The imprints of the disc build-up scenarios}

Considering suitable scenarios for the \N7572 disc formation, first of all we stress that its thick disc could not be created from stars of accreted dwarf galaxies, because normally, stellar populations in dwarfs have lower metallicities and [Mg/Fe] ratios \citep[see e.g.][]{Chilingarian2009} than what we observe in \N7572.
According to \citet{Forbes2004}, stellar population properties of the MW thick disc do not agree with such a model either.

\subsubsection{Kinematics of stellar disc components}

Observational data indicate that stellar populations of the thin and thick discs of the MW have different kinematics, in particular the velocity dispersions and the mean orbital radii \citep{Lee2011,Yu2018,Mackereth2019}.
Several numerical models challenge the ability of radial migration alone to create a thick disc from stars with large dispersion by moving stars born in the inner regions of the disc away from the mid-plane \citep[see][]{Minchev2012}. 
However, \citet{Loebman2011} and \citet{RoskarDebattista+2013} have reproduced this effect in their models. 
In case of \N7572, the inner region of the disc/bulge has a velocity dispersion value comparable to that of a thick disc. 
The observed very old ages in this area suggest that the later star formation episode did not significantly affect that region. 
The high stellar metallicity and relatively low [Mg/Fe] exclude the possibility that the inner region in the mid-plane is the source of stars observed in the thick disc of \N7572.

Although we do not have the data to build the AVRs for \N7572 for the analysis of the thin disc heating process, some observed features allow us to reject the thin disc heating scenario.
The velocity dispersion of the \N7572 thin disc is greater than that of the MW thick disc \citep{Bland-Hawthorn2016}, which is inert and not subject to additional heating.
Moreover, if a thick component is the result of a thin disc heating via the influence of giant molecular clouds, then the radial velocity dispersion of the high--[$\alpha$/Fe] component would be lower than observed. 
Indeed this heating mechanism is not sufficient to explain the observed properties of the MW thick disc \citep{Aumer2016,Yu2018}. 
Unless GMCs were much more effective at scattering stars in \N7572 than in the MW, the same must hold true in \N7572.  
Hence, it is more natural to explain the observed ratio of velocity dispersions of \N7572 discs by having its thick disc formed before the thin disc.

\subsubsection{The radial scalelengths of the disc components}

The ratio of the radial scalelengths of the thin and thick components is an important indicator of the disc formation scenario.
In upside-down scenarios, the radial scale of a younger thin disc should be larger than that of a thick disc \citep[see for example ][]{Bird2013, Lehnert2014} because a thin disc is formed from gas having a higher angular momentum than the thick disc material.  
According to the models by \citet{Stinson2013} the scalelengths of the stellar populations with narrow [Fe/H] and [$\alpha$/Fe] abundances ranges grow non-linearly with age.
In a MW-like galaxy we will be able to notice an increase in the radial scale of the disc component with age from observations only for a population younger than 5~Gyr because all older populations have the same scalelengths \citep[see cosmological hydro-dynamical simulations by][Fig.~8]{Stinson2013}.

However, comparing these models with observations is difficult even for the MW because it is not always possible to unambiguously separate  thick and thin disc stellar populations.
For example, in the Solar neighbourhood stars with a high [Mg/Fe] content  likely have diverse origins, not only they come from the thick disc but also from the inner regions of the thin disc \citep{Haywood2013,Hayden2017}. 
Nevertheless, according to several recent studies \citep[e. g.][]{Carollo2010,Cheng2012,Bovy2016,Mateu2018}, the radial scale of the MW thick component having a high [$\alpha$/Fe] ratio is similar or slightly shorter than that of the thin disc, which is approximately $2.6$~kpc \citep{Bland-Hawthorn2016}. 
On the contrary, \citet{Lopez-Corredoira2014} using simple star counts conclude that the thick disc radial scale is longer than that of the thin component.

In \N7572, the equal values of the mid-plane radial scales of the disc components with ages $5\dots8$~Gyr and $10\dots13$~Gyr are in good agreement with the upside-down formation scenario. 
However, if we integrate the vertical distributions of the disc components at every $R$, the scalelength of the resulting radial distribution of the thick disc will become longer than that of the thin disc (see Sect.~\ref{T2t_contribution}).

In other (edge-on) galaxies, scales of components are defined using photometric decomposition techniques, which makes it difficult to compare them with the results for the MW.  
According to \citet{Comeron2012}, the thick discs usually have larger scalelengths than thin discs \citep[see also][]{Pohlen2004,YoachimDalcanton2006}.
These values apply both to the mid-plane and to any $z$ because the flaring is usually neglected. 
But, as we demonstrated above, the radial scales can be greatly overestimated in case of significant disc flaring and also due to other purely geometric effects.

\subsubsection{Disc flaring}

In recent photometric studies of the vertical structure of disc galaxies \citep[][and many others]{YoachimDalcanton2006,Comeron18}, the thickness variations of both disc components with radius were neglected.
This assumption relied on earlier observational studies \citep[see e. g.][]{vdKS81,Bizyaev2002}, even though there are hints that vertical scaleheights are constants with radius only in late-type galaxies (see \citet{deGrijsPeletier1997} and fig.~9 in \citet{Bizyaev2014}).
\citet{Pinna2019_1,Pinna2019_2} demonstrated significant thin disc flaring in two out of three studied S0 galaxies in the Fornax cluster. 
The thin disc of the MW broadens in the outskirts while there is still a vigorous discussion about the thick disc  \citep{Lopez-Corredoira2014,Bland-Hawthorn2016, Bovy2016}.
However, as we demonstrated earlier, in \N7572 we cannot ignore the change of the thicknesses of both disc components, even though because of our simplified mathematical approach, the exact form of the component broadening cannot yet be precisely calculated.

If a thick disc forms in an \textit{isolated} galaxy at high redshift (z~$\sim 2$) in the conditions of high-velocity dispersion of the interstellar medium \citep{Elmegreen2006, Bournaud2009,Lehnert2014}, a dynamically hot thick disc would have a constant vertical scale. 
Afterwards it becomes insensitive to additional heating.
But if some mergers occurred along the rapid formation stage of a thick disc, according to  \citet{Bournaud2009} and \citet{Qu2011} the resulting thick stellar component would flare. 
Because \N7572 is an overmassive giant galaxy in a dense environment, having a number of minor mergers in its early formation history looks quite plausible and it will explain the significant flaring of the thick disc that we observe.

\subsubsection{The mass ratio of the disc components}

The mass ratio of the thick to thin disc components is thought to anticorrelate with the total mass of the system \citep{YoachimDalcanton2006,Comeron18}. 
It is about $1/3$ for massive galaxies but in intermediate-mass systems with rotation velocities $v_{\mathrm{rot}}<120$~{\kms} the thick and thin component masses can be equal.
We note that to correctly estimate the thick and thin disc masses of external galaxies, in addition to accurate photometric decomposition, it is necessary to know the mass-to-light ratios of the components.
The use of the ``mass follows light'' assumption in the optical bands is a too big of a simplification, because as we demonstrated in \citet{Kasparova16} and \citet{Katkov2019}, stellar population properties of thick and thin discs can differ significantly \citep[see also][]{Comeron+2015,Pinna2019_2}.

In \N7572 the mass ratio of the thick and thin disc components is $M_T/M_t\approx2.7$ and it is an unprecedentedly large value for such a massive galaxy.
Nonetheless, we can simply explain this fact within the scenario of the rapid formation of its thick disc and subsequent gradual growth of the thin disc. 
In a two-stage scenario there must be objects whose thin discs have not yet grown or became underdeveloped because of environmental effects. 
The latter situation can happen if a cold gas reservoir (e.g. intergalactic filaments or settling gas ejected by galactic winds) had been stripped or exhausted before the thin disc grew to the ``normal'' size. 
The proposed formation mechanism of \N7572 is in a good agreement with a scenario by \citet{Silchenko2012} explaining the general domination of lenticular galaxies over spirals in dense environment by the fact that they did not have enough material to form a thin disc. 
This scenario can be realized due to a number of effects associated with a dense environment, e.g ram pressure stripping \citep{GunnGott1972}, interaction with nearby galaxies \citep{Mihos2004}, starvation \citep{Larson+80} and others. 
\citet{Comeron16} and \citet{Katkov2019} also showed the significant influence of the environment on the thin disc growth in several cluster galaxies. 

We believe that our observational data analysis for \N7572 provides an example on a disc formation mechanism similar to for the MW that acts in the extreme case of a dense cluster environment for a very massive giant galaxy.

\subsection{How a massive thick disc can affect the~interpretation of galaxy observations}

The example of \N7572 clearly shows that it is naive to assume that for any non-edge-on oriented galaxy most of its light comes from a thin component of the disc. 
When we study galaxies with arbitrary orientation of their disc plane in the sky we must keep in mind the multicomponent nature of their discs. 
This can significantly complicate the analysis of stellar population properties (especially if they are old enough) in cases of non-edge-on disc galaxies because their thick disc components will affect both kinematics and stellar population measurements. 
In certain regions of galaxies we will see a mixture of stellar populations from layers formed in very different conditions at different redshift and star formation histories. 
If we turn \N7572 to the face-on projection, more than 60~per~cent of the light will come from the old hot thick disc and its contribution will grow to the outskirts.
Then we will underestimate its rotation velocity significantly due to the large velocity dispersion of the thick disc and, in fact, it might be classified as a rotating elliptical galaxy.
In some cases, observed gradients of stellar population properties can be the result of projection effects of two sub-populations on the line-of-sight. 
There are similar observational manifestations along the antitruncation features in surface brightness profiles if the flaring is insignificant but the radial scale of the thick disc in the mid-plane is larger than that of the thin disc  \citep{Comeron2012}. 
These effects combined with a possible thin disc truncation allow us to study stellar populations of a thick disc not only in the mid-plane of edge-on galaxies \citep{Katkov2019} but also in the periphery of discs at any orientation to the line-of-sight.

\section{Summary}
\label{summary}

In this study we first present a novel data analysis techniques and then apply them to deep imaging and spectroscopic observational data for the extremely luminous and large lenticular galaxy \N7572.

Using simple 1D decomposition of vertical cross-sections of the disc and assuming a double-$\sech^2$ model we detected and measured changes in the vertical scaleheights of both thick and thin discs and measured their radial scalelengths.  
We also obtained a map of the relative contributions of a thick and thin disc required for a physically motivated interpretation of spectral data where we observe the overlap of stellar populations in the line of sight.
We demonstrated the importance of taking into account the individual contributions of the thick and thin discs to the mid-plane light profile required to correctly estimate radial scales of the both components.  
We showed that the thick disc radial scale as obtained by fitting the surface brightness profiles above the mid-plane can be overestimated because of the flaring and imperfect edge-on disc orientation.

As a part of the analysis of mid-plane spectra using realistic assumptions about stellar dynamics, we reconstructed the stellar LOSVD in a non-parametric way and fitted it using a two-component parametric model that took into account the orbital distribution of stars on the line of sight.
This is possible for \N7572 because of the significant difference in the kinematics of its thick and thin discs.

We carried out the first spectrophotometric study of the vertical structure of a giant disc galaxy significantly exceeding the size of the Milky Way.
\N7572 has a total disc mass of $\sim  2.2 \times 10^{11}$~\Ms{}, a circular rotation velocity of $\sim370$~{\kms} and it lives in a dense environment.

Our photometric analysis of DECaLS-$g$ images demonstrates that:
\begin{itemize}
\item We need at least two components to describe the vertical structure of the disc in \N7572;
\item Both disc components exhibit flaring. 
The scaleheight of the thin disc in the inner part is $z_t=1.5\pm0.1$~kpc and it grows to $\sim2$~kpc in the outer regions.
The scaleheight of the thick component grows from $z_T=6.0\pm0.2$~kpc by a factor of two;
\item The scalelengths of the thin and thick components are almost identical ($h_t = 5.4\pm0.1$~kpc and $h_T = 5.6\pm0.3$~kpc);
\item The thick disc contribution to the disc brightness changes from 30~per~cent to 90~per~cent moving from the mid-plane to $|z|=4$~kpc above it and it only slightly changes with the radius;
\item We found some excess flux noticeable at $|z| > 15$~arcsec, not described by the former two components, which can be either (i) the sign of the formation process of a thick disc through minor mergers or (ii) the effect of the line-of-sight integration of the light from two not exactly sech$^2$--like flaring components or (iii) an additional third disc component.
\end{itemize}

We analysed deep spectroscopic long-slit observations collected with the Russian 6-m BTA telescope.
We obtained spectra at two positions, the major axis (mid-plane) and the region parallel to the major axis offset by 5~arcsec ($\approx4.0$~kpc) above the mid-plane.
We do not see statistically significant radial gradients of the stellar population properties (age, [Fe/H], [Mg/Fe]). 
The thick disc has an older and dynamically hotter stellar population than the thin disc, the $\alpha$-element abundances [Mg/Fe] correspond to $+0.15$ and $+0.3$~dex for the thin and thick  components respectively.
The non-parametric stellar LOSVD along the mid-plane shows that the thick disc rotates 200~\kms\ slower than the thin disc.
Despite significant differences in mass, scalelengths and kinematics, mean ages and metallicities of the disc components are close to those for ``ordinary'' giant early-type disc galaxies: $T_{{\rm SSP},t} = 5\dots8$~Gyr and $T_{{\rm SSP},T}=10\dots13$~Gyr,  [Fe/H]$_{{\rm SSP},t}=0.0$~dex and [Fe/H]$_{{\rm SSP},T}=-0.1$~dex for the thin and thick components respectively.

We derived the stellar mass-to-light ratios $(M/L)_{t}=3.1\, (M/L)_{\odot}$ and $M/L_{T}=4.6\, (M/L)_{\odot}$ (SDSS $g$-band) for the thin and thick discs and estimated their stellar masses to be $M_t=5.9\times10^{10}$~\Ms\ and $M_T=1.6\times10^{11}$~\Ms.
Hence, the thick-to-thin mass ratio is $M_T/M_t=2.7$.

Our results allow us to draw a number of important conclusions regarding the analysis of observational data of non-edge-on oriented galaxies, at least for massive giant discs. 
It is always necessary to take into account the multicomponent nature of discs due to a potential significant contribution of a thick disc to the total light. 
In some regions of galaxies we will see a mixture of stellar populations from layers formed in very different conditions at different epochs.
Because of the flaring, ``face-on'' radial scales can differ from the values in the mid-plane (in the case of \N7572 it is $6.6 \pm 0.4$~kpc and $9.1 \pm 1.8$~kpc for the thin and thick components) which complicates the analysis of stellar population gradients. 

At the end, our analysis allowed us to choose the most plausible scenario explaining the observed vertical structure of \N7572.
At high redshift (z $\sim2$) the progenitor of a giant galaxy in the process of a wild star-forming phase suffers by some minor mergers and gives birth to a thick giant flaring disc with large radial and vertical scales.
Then, after some time, the thin disc formation starts because of either additional gas accretion from infalling gas-rich satellites or accreted through filaments or the cooling of the gas remained after the phase of the thick disc formation.
For some reason about 5~Gyr ago the cold gas reservoir was exhausted or removed earlier than in most galaxies, e.g. due to the influence of a dense cluster environment. 
This stopped the thin disc formation in \N7572 prematurely so that it could not grow further.

\section*{Acknowledgments}

We are grateful to Anatoly Zasov and Olga Sil'chenko at Sternberg Astronomical Institute, Moscow State University for useful suggestions and fruitful discussions. We thank Alexey Moiseev and Dmitri Oparin at Special Astrophysical Observatory, Russian Academy of Sciences for their assistance in our SCORPIO observations.
We are thankful to the anonymous referee for the important comments.
The development of advanced data analysis techniques for deep spectroscopic observations is supported by the Russian Science Foundation project 19-12-00281.
The authors acknowledge the support from the Program of development of M.V. Lomonosov Moscow State University for the Leading Scientific School ``Physics of stars, relativistic objects and galaxies''.

The 6-m telescope of the Special Astrophysical Observatory of the Russian Academy of Sciences is operated with the financial support of the Ministry of Science and Higher Education of the Russian Federation. 

This research has made use of the Lyon Extragalactic Database (LEDA, http://leda.univ-lyon1.fr/).

This research has made use of the NASA/IPAC Extragalactic Database, which is funded by the National Aeronautics and Space Administration and operated by the California Institute of Technology.

In this study we used the data from the Dark Energy Camera Legacy Survey (DECaLS; NOAO Proposal ID \#2014B-0404; PIs: David Schlegel and Arjun Dey) available online at \url{http://legacysurvey.org} obtained at the Blanco telescope, Cerro Tololo Inter-American Observatory, National Optical Astronomy Observatory (NOAO) with the Dark Energy Camera (DECam), which was constructed by the Dark Energy Survey (DES) collaboration. The Legacy Surveys imaging of the DESI footprint is supported by the Director, Office of Science, Office of High Energy Physics of the U.S. Department of Energy under Contract No. DE-AC02-05CH1123, by the National Energy Research Scientific Computing Center, a DOE Office of Science User Facility under the same contract; and by the U.S. National Science Foundation, Division of Astronomical Sciences under Contract No. AST-0950945 to NOAO.

%%%%%%%%%%%%%%%%%%%%%%%%%%%%%%%%%%%%%%%%%%%%%%%%%%


\bibliographystyle{mnras}
\bibliography{Thicks_disk_art}

%%%%%%%%%%%%%%%%%%%%%%%%%%%%%%%%%%%%%%%%%%%%%%%%%%

%%%%%%%%%%%%%%%%% APPENDICES %%%%%%%%%%%%%%%%%%%%%

\appendix

\section{Testing the photometric decomposition technique} \label{appendix}

\begin{figure*}

\includegraphics[scale=0.8,trim={1.0cm 0cm 0cm 0cm}]{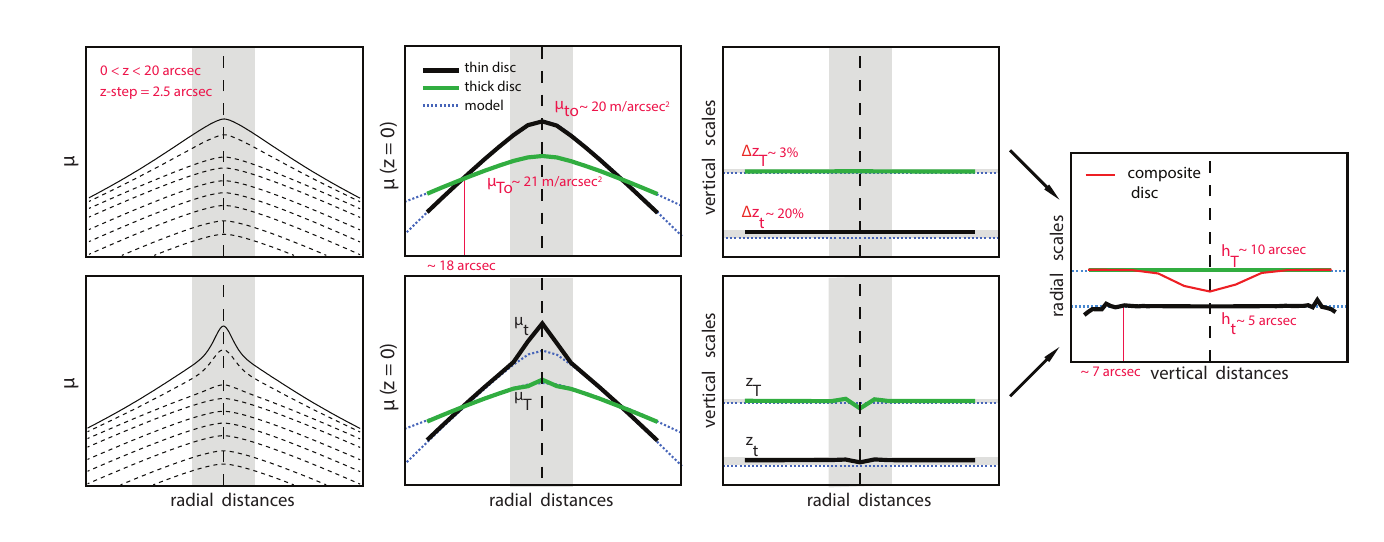}
\caption{The results of the reconstruction of photometric parameters for our Model~1 (upper row) and Model~2 (bottom row).
From left to right: radial profiles at different $z$, the mid-plane surface brightness, vertical scales of the components, radial scales restored from the vertical profile decomposition (almost identical for both models).
The black and green thick curves correspond to a thin disc and thick discs respectively.
The blue dotted lines show the disc parameters of the input models.
The red thin line in the right-most panel shows the estimates of the radial scales of the composite disc (thin+thick) at different $z$.
The red numbers correspond to the approximate values for the 2D model of the disc of \N7572 with constant thicknesses of both components.
\label{tests}}
\end{figure*}

\begin{figure}

\includegraphics[scale=0.59]{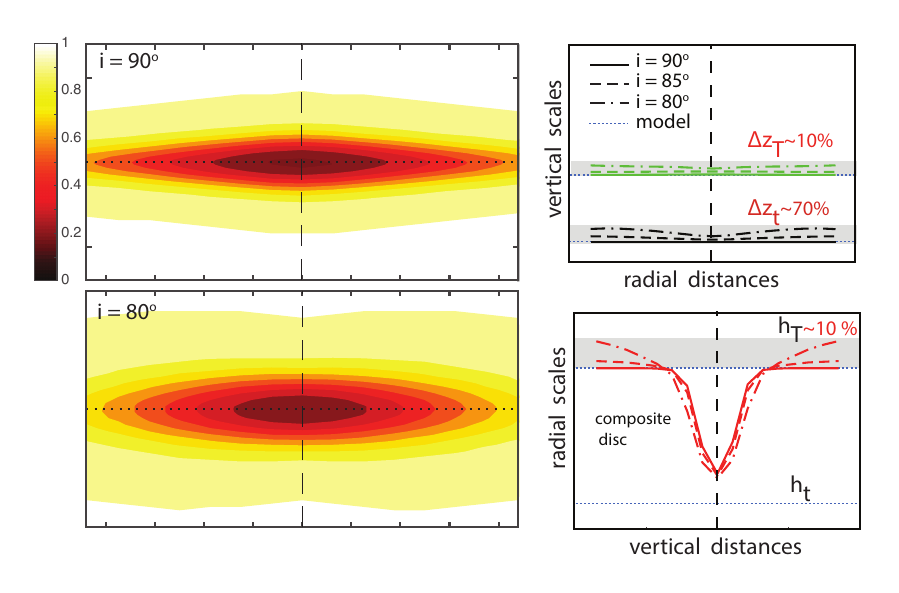}
\caption{Left-hand panels: the maps of the intensity ratio of the thick disc to the total disc (thick $+$ thin components) in cases of Model~1 with $i=90$~deg (top) and $i=80$~deg (bottom).
Right-hand panels: the output vertical scales (top) and the profile of the composite radial scales (bottom).
\label{tests_fig2}}
\end{figure}

We made a series of tests to predict the behaviour of our simplified method for estimating photometric parameters.
We applied it to a model galaxy with a disc having two components, and where the radial scale of the thick disc is larger than that of the thin disc. 

For this purpose we created two toy models of \N7572: a model with a thin and a thick disc (Model~1) and a model with two discs and a bulge (Model~2).
Both models are convolved with the DECam PSF. 
All discs are constructed using Eq.~\ref{eq_sech2} and Eq.~\ref{eq_exp_edge-on} \citep{vdKS81} and they have the constant vertical scales $z_t(r)=$~const and $z_T(r)=$~const, the radial scales of the components are set to be $h_T > h_t$.
The parameters are set to be close to the values for \N7572. 
In the case of Model~2 we used S{\'e}rsic bulge\footnote{The parameters of the S{\'e}rsic bulge are $I_0=2.4\times10^3$~\Ls~pc$^{-2}$, $R_e = 984$~pc, $n=0.8$.}. 
The composite radial light profiles (thin $+$ thick) in both cases do not have any notable breaks outside the bulge area.

We show the results in the form of a diagram in Fig.~\ref{tests} (top row for Model~1, bottom row for Model~2).
The black and green lines indicate the values of the vertical scales restored by our method, the brightness of the disc components in the mid-plane, and their radial scales. 
The blue  dotted lines show disc parameters of the input models.
To estimate radial scales, we fit the radial profiles of each component excluding the bulge region, thus the last panel in the rows is identical for Models~1 and 2.

Our test demonstrates that:
\begin{enumerate}
\item  It is sufficient to exclude the area affected by the bulge ($\pm8$~arcsec in case of \N7572) to obtain unbiased estimates of the disc scales and it is not required to include an analytic description of the bulge in the equations.
Estimates of the vertical scales outside the bulge area coincide with the input values in the model.
\item The vertical scale of the thin disc is overestimated by 20~per~cent while for the thick disc only by 3~per~cent \textit{only} due to the influence of the PSF.
The estimates of the radial scales and the mid-plane  surface brightnesses do not change much. 
\item The restored scaleheights do not change with radius (as in the input models) even at the periphery of the galaxy where the surface brightnesses of the model thick and thin discs are compared.
\item The radial scale evaluation of the composite disc $h$ does not correspond to the input thin disc radial scale even in the mid-plane because the contribution of the thick disc makes the brightness profile of a galaxy significantly flatter.
However, when we go far enough above the mid-plane, the radial scale reaches a plateau $h \longrightarrow h_{T}$.
\item Our method for reconstructing the radial scales of both components from the vertical decomposition works correctly over wide range of $z$ values, except for the regions placed high above the mid-plane (which for Models~1 and 2 corresponds $z\approx 7$~arcsec and $\mu_{z = 0} \approx 25$~mag arcsec$^{-2}$) where the thin disc contribution is negligible. 
\end{enumerate}

We performed the second series of tests because we do not know how accurately the inclination angle of the object is determined. 
\citet{Comeron11} and \citet{Devour2017} demonstrated the importance of this issue. 
Actually, in late-type disc galaxies, the dust layer position helps us. 
However for lenticular galaxies without well-defined dust lanes there is no easy diagnostics for the disc orientation.
\citet{ZasovKhoperskov2003} showed that the effect of the integration along the line-of-sight for edge-on galaxies causes that the LOS velocity gradually reaches the real circular velocity only at a distance of several radial scales. 
That is, according to \citet{ZasovKhoperskov2003}, we suspect that the galaxy is not precisely edge-on oriented because we see a pronounced plateau of the line-of-sight (LOS) velocity distribution in its mid-plane. 
We would like to test how well we can estimate the parameters when a studied galaxy is not seen exactly edge-on.
To carry out this test, we created three models, similar to Model~1, described above, but with three inclinations angles $i = 80$, 85 and 90~deg.

We presents the results of this series of tests in Fig.~\ref{tests_fig2}:
\begin{enumerate}
\item The first result is quite expected: when decreasing $i$, the region of the thin disc dominance becomes more compact in $R$ and thicker in $z$.
In the mid-plane the contribution of the thick disc will increase at all distances from the centre in comparison of the edge-on case.
From the point of view of the observation analysis, it means that we will see a smaller difference between the stellar population properties in the mid-plane and at large $z$--distances. 
\item If $i$ differs from 90 deg, then the composite radial scale does not approach a plateau with $h=h_T$. 
It continues to grow above $h_T$ with the increasing distance from the mid-plane. 
However even for the model with $i=80$ deg, this effect is within the photometrical measurement errors and it is much smaller than the influence of the disc flaring on the scalelength estimates.
\item The smaller $i$, the more overestimated values we get for the thicknesses of both components of the disc.
In our case, the height of a thick disc can be overestimated by 10~per~cent, and for the thin disc by 70~per~cent if the inclination is off by 10~deg. 
If $i=85$~deg only $\Delta z_t$ is substantial ($\sim 30$~per~cent).
\citet{Comeron11} obtained similar results by testing their model in case of galaxies seen not perfectly edge-on.
Both scaleheight estimates increase for $i < 90$~deg, but for a thin disc it is much more significant (see fig.~9 in their work).
\item 
If the radial scales of the disc components are different, the observed gradients of the stellar population properties in the mid-plane do not reflect the true features of the thin disc, but the change in the mutual contribution of the components. 
These gradients will change significantly when $i$ departs from 90 deg.

\end{enumerate}

%%%%%%%%%%%%%%%%%%%%%%%%%%%%%%%%%%%%%%%%%%%%%%%%%%

% Don't change these lines
\bsp	% typesetting comment
\label{lastpage}
\end{document}